
%
%
\input harvmac
%
%
%
%
\ifx\answ\bigans
\else
\output={
  \almostshipout{\leftline{\vbox{\pagebody\makefootline}}}\advancepageno
}
\fi
%
%
%
\def\mayer{\vbox{\sl\centerline{Department of Physics 0319}%
\centerline{University of California, San Diego}
\centerline{9500 Gilman Drive}
\centerline{La Jolla, CA 92093-0319}}}
%
%
\def\doe{\#DOE-FG03-90ER40546}
\def\pyiam{PHY-8958081}

%
%
\def\UCSD#1#2{\noindent#1\hfill #2%
\bigskip\supereject\global\hsize=\hsbody%
\footline={\hss\tenrm\folio\hss}}
%
%
\def\abstract#1{\centerline{\bf Abstract}\nobreak\medskip\nobreak\par #1}
%
%
%
%
\edef\tfontsize{ scaled\magstep3}
 \tfontsize  \tfontsize
 \tfontsize \font\titlei=cmmi10 \tfontsize
\font\titleis=cmmi7 \tfontsize \font\titleiss=cmmi5 \tfontsize
\font\titlesy=cmsy10 \tfontsize \font\titlesys=cmsy7 \tfontsize
\font\titlesyss=cmsy5 \tfontsize  \tfontsize
\skewchar\titlei='177 \skewchar\titleis='177 \skewchar\titleiss='177
\skewchar\titlesy='60 \skewchar\titlesys='60 \skewchar\titlesyss='60
%
%
%
%
%
\def\inv{^{\raise.15ex\hbox{${\scriptscriptstyle -}$}\kern-.05em 1}}
\def\lbar{{\lower.35ex\hbox{$\mathchar'26$}\mkern-10mu\lambda}} 

%
%
%
%
\def\slash#1{\rlap{$#1$}/} 
\def\dsl{\,\raise.15ex\hbox{/}\mkern-13.5mu D} 
\def\delsl{\raise.15ex\hbox{/}\kern-.57em\partial}
\def\Ksl{\hbox{/\kern-.6000em\rm K}}
\def\Asl{\hbox{/\kern-.6500em \rm A}}
\def\Dsl{\hbox{/\kern-.6000em\rm D}} 
\def\Qsl{\hbox{/\kern-.6000em\rm Q}}
\def\gradsl{\hbox{/\kern-.6500em$\nabla$}}
%
%
\def\lspace{\ifx\answ\bigans{}\else\qquad\fi}
\def\lbspace{\ifx\answ\bigans{}\else\hskip-.2in\fi} 
%
%
\def\boxeqn#1{\vcenter{\vbox{\hrule\hbox{\vrule\kern3pt\vbox{\kern3pt
        \hbox{${\displaystyle #1}$}\kern3pt}\kern3pt\vrule}\hrule}}}
%
%
\def\mbox#1#2{\vcenter{\hrule \hbox{\vrule height#2in
\kern#1in \vrule} \hrule}}
%
%
%
%
   
   \def\CH{{\cal H}}
   \def\CL{{\cal L}}

%
%
%
%
%

%

\def\bar#1{\overline{#1}}

\def\ket#1{\left| #1\right\rangle}
\def\abs#1{\left| #1\right|}

\def\darr#1{\raise1.5ex\hbox{$\leftrightarrow$}\mkern-16.5mu #1}

%
%
\def\frac#1#2{{\textstyle{#1\over #2}}} 
%
%
%
%

\def\Tr{\mathop{\rm Tr}}

\def\MeV{{\rm MeV}}

%
%
%
%

%
%
\def\ltap{\ \raise.3ex\hbox{$<$\kern-.75em\lower1ex\hbox{$\sim$}}\ }
\def\gtap{\ \raise.3ex\hbox{$>$\kern-.75em\lower1ex\hbox{$\sim$}}\ }
\def\gl{\ \raise.5ex\hbox{$>$}\kern-.8em\lower.5ex\hbox{$<$}\ }
\def\roughly#1{\raise.3ex\hbox{$#1$\kern-.75em\lower1ex\hbox{$\sim$}}}
%
%
        \def\etc{\hbox{\it etc.}}
\def\eg{\hbox{\it e.g.}}

\def\np#1#2#3{Nucl. Phys. B{#1} (#2) #3}
\def\pl#1#2#3{Phys. Lett. {#1}B (#2) #3}

\def\physrev#1#2#3{Phys. Rev. {#1} (#2) #3}

\relax

\noblackbox
\def\twolr{SU(2)_L\times SU(2)_R}
\def\split{$\Sigma_Q^*-\Sigma_Q$}
\def\kket#1{\left.\left|#1\right\rangle\!\right\rangle}
\def\twobox#1#2{\vcenter{\hrule \hbox{\vrule height#2in
\kern#1in \vrule} \hrule \hbox{\vrule height#2in
\kern#1in \vrule}\hrule}}
\def\anti{\,\twobox{.1}{.1}\,}
\def\fund{\,\mbox{.1}{.1}\,}
\def\fundbar{\,\bar{\mbox{.1}{.1}}\,}
\def\bfrac#1#2{{{#1}\over{#2}}}
\def\ssq{\Sigma^*_Q}
\def\sq{\Sigma_Q}
\def\pq{P_Q}
\def\psq{P^*_Q}
\def\lq{\Lambda_Q}

\centerline{{\titlefont{Hyperfine Splittings of
Baryons Containing a}}}
\medskip
\centerline{{\titlefont{Heavy Quark in the Skyrme Model}}}
\bigskip
\centerline{Elizabeth Jenkins and Aneesh V. Manohar}
\bigskip
\mayer
\vfill
\abstract{The $\Sigma_c^*-\Sigma_c$ and $\Sigma_b^*-\Sigma_b$ hyperfine
mass splittings are computed in the Skyrme model. The hyperfine
splittings are suppressed by both $1/N_c$ and by $1/m_Q$, where $N_c$
is the number of colors and $m_Q$
is the mass of the heavy quark. The $\Sigma_c$, $\Sigma_c^*$, $\Sigma_b$,
$\Sigma_b^*$, and $\Lambda_b$ masses are predicted in terms of the known
values of the $\Lambda_c$, $D$, $D^*$, $B$ and $B^*$ masses.
}
\vfill
\UCSD{\vbox{\hbox{UCSD/PTH 92-26}\hbox{hep-ph/9208238}}}{August 1992}

Baryons containing a heavy quark can be interpreted as bound states of
mesons containing a heavy quark and chiral solitons of the $\twolr$
nonlinear sigma model. This approach was originally proposed by Callan and
Klebanov to treat baryons containing an $s$ quark as soliton--$K$-meson
bound states \ref\calkleb{C. Callan and I. Klebanov, \np{262}{1985}{365}\semi
C. Callan, K. Hornbostel and I. Klebanov \pl{202}{1988}{269}}.
Early attempts were made to extend this method
to baryons containing a heavy
quark $Q$, such as the $c$ or $b$ quark \calkleb\ref\mrho{M. Rho, D.O.
Riska, and N.N. Scoccola, \pl{251}{1990}{597},
Z.~Phys. {A341} (1992) 343\semi
Y. Oh, D. Min, M. Rho, and N. Scoccola, Nucl. Phys. A534 (1991)
493}.  A correct treatment of heavy baryons as soliton--heavy-meson
bound states must incorporate the consequences of heavy quark
symmetry \ref\jmw{E. Jenkins, A.V.
Manohar and M.B. Wise, Caltech Preprint CALT-68-1783 (1992)}.
In the heavy quark limit, the $D$ and $D^*$ ($B$ and $B^*$) mesons
are degenerate, so the effective chiral theory must include both the
pseudoscalar ($\pq$) and vector meson ($\psq$) fields in order to
respect the heavy quark symmetry.  Properties of the heavy baryon
bound states have been calculated to leading order in $1/ m_Q$
\jmw\ref\glm{Z. Guralnik, M. Luke and A.V. Manohar,
UCSD/PTH 92-24 (1992)}.
The Skyrme model successfully predicts the existence of $\Lambda_Q$, $\sq$
and $\ssq$ bound states \jmw\ and leads to mass relations which are
well-satisfied by the measured $\Lambda_{c,b}$ and $\Sigma_{c,b}$ masses \glm.
In the heavy quark limit, the $\sq$ and $\ssq$ are a degenerate
multiplet with isospin one, spin of the
light degrees of freedom one, and total spin 1/2 and 3/2, respectively.
The energy splitting between these states is a $1/m_Q$ effect.  In this
letter, the $\ssq -\sq$ mass difference is computed.

It is instructive to first study the hyperfine splittings in
the constituent quark model. The \split\ and $P_Q^*-P_Q$
mass differences are due to
the hyperfine interaction generated by one-gluon exchange between
constituent quarks \ref\dgg{A. De Rujula, H. Georgi
and S.L. Glashow, \physrev{D12}{1975}{147}},
\eqn\hyper{
\CH = -{\kappa\over N_c}
 \sum_{<ij>} T^A_i T^A_j\ {S_i\cdot S_j\over m_i m_j},
}
where $T^A_i$, $S_i$ and $m_i$ are
the color generator, spin and mass of the $\imath^{\rm th}$ quark, $N_c$
is the number of colors, and
the sum is over all pairs. The constant $\kappa$ depends on the
wave function of the hadron, and may be different for mesons and
baryons.
$\CH$ has an overall factor of
$1/N_c$, since the gluon coupling constant is of order
$1/\sqrt{N_c}$.
The pseudoscalar and vector mesons $P_Q$ and $P^*_Q$ are
bound states of a heavy quark $Q$ and a light antiquark $\bar q$
in a color singlet state, with spin zero and one, respectively.
The
hyperfine interaction Eq.~\hyper\ in the meson sector can be
written in the form
\eqn\mhyperone{
\CH = -{\kappa\over 4 m_Q m_q N_c}
 \left[(T_Q+ T_{\bar q})^2
- T_Q^2 - T_{\bar q}^2\right] \left[
(S_Q+S_{\bar q})^2 - S_Q^2 - S_{\bar q}^2\right].
}
For the vector meson $\psq$, $(S_Q+S_{\bar q})^2=2$, and for the
pseudoscalar meson $\pq$, $(S_Q+S_{\bar q})^2=0$. The color factors
can be written in terms of
quadratic Casimirs defined by
\eqn\casimir{
(T_R^A\ T_R^A)^i_j = c(R)\ \delta^i_j,
}
for the $SU(N_c)$ representation $R$.
This gives the hyperfine splitting
\eqn\mhypertwo{
\psq-\pq =  {\kappa\over 2 m_Q m_q N_c}
\left[c(\fund)+c(\fundbar)-c(1)\right],
}
where the irreducible representations of $SU(N_c)$ are denoted by Young
tableaux.

The hyperfine interaction
Eq.~\hyper\ in the baryon sector can be written in the form
\eqn\mbaryone{
\CH = {\kappa'\over 2 N_c}
 \left[c(\fund)+c(\fund)-c({\anti})\right] \left[
\sum_{i=1}^{N_c-1} {S_Q\cdot S_i\over m_Q m_q}
+ \sum_{<ij>}{S_i\cdot S_j\over m_q^2}\right],
}
where the first term is the interaction of the heavy quark with the
light quarks, and the second term is summed over all light quark pairs.
The last Casimir is that of the two-index antisymmetric tensor, because
any two quarks in a baryon are antisymmetric in color.
The spin of the light degrees of freedom is the spin of the light quarks,
$S_\ell=\sum_{i=1}^{N_c-1}S_i$, so
that $\CH$ is
\eqn\mbarytwo{
\CH = {\kappa'\over 2 N_c}
 \left[2c(\fund)-c({\anti})\right]\left\{ \left[
{S_Q\cdot S_\ell\over m_Q m_q} \right]
+ \CH_\ell\right\},
}
where $\CH_\ell$ is the interaction energy of the light degrees of
freedom. $\CH_\ell$ is of order one, since all the explicit $N_c$
dependence has been factored out.
The $\Sigma_Q$ and $\Sigma^*_Q$ have light degrees of freedom
in identical quantum states so that $\CH_\ell$ cancels in
the mass difference. Using $2 S_Q\cdot S_\ell = (S_Q+S_\ell)^2 - S_Q^2
-S_\ell^2$, with $s_Q=1/2$, $s_\ell=1$, and
$s_Q+s_\ell=1/2$ for the $\sq$ and
$s_Q+s_\ell=3/2$ for the $\ssq$, gives
\eqn\mbarythree{
\Sigma_Q^*-\Sigma_Q ={3\kappa'\over 4 m_Q m_q N_c }
\left[2c(\fund)-c({\anti})\right].
}
The $\lq$ state is a spin-1/2 baryon with the spin of the light degrees
of freedom $s_\ell=0$, so the first term in Eq.~\mbarytwo\ does not
contribute to the $\lq$ mass. The first term in Eq.~\mbarytwo\ also
cancels out of the average mass of the $\sq,\, \ssq$
multiplet $(\sq)_{\rm avg}= \frac 1 3 (\sq + 2 \ssq)$.
The $(\sq)_{\rm avg}-\lq$ mass difference is
\eqn\sldiff{
(\sq)_{\rm avg}-\lq  = {\kappa'\over 2 N_c}
 \left[2c(\fund)-c({\anti})\right]  \Delta \CH_\ell,
}
where $\Delta \CH_\ell$ is the difference in interaction energies of the
light degrees of freedom in the $\lq$ and $\sq,\, \ssq$.

The Casimirs are trivial to compute using standard techniques of group
theory \ref\georgi{H. Georgi, Lie Algebras in Particle Physics,
(Benjamin/Cummings, 1982), Problem XVI.C},
\eqn\casvalues{
c(1)=0,\ \  c(\fund)=c(\fundbar)={N_c^2-1\over2N_c},\ \  c({\anti})={
(N_c-2)(N_c+1)\over N_c}.
}
The color factor for the mesons is
\eqn\mcolor{
c(\fund)+c(\fundbar)-c(1)= {N_c^2-1\over
N_c}\rightarrow  N_c \quad({N_c\rightarrow\infty}),
}
and for the baryons is
\eqn\bcolor{
2c(\fund)-c({\anti})= {N_c+1\over N_c}
\rightarrow  1\quad ({N_c\rightarrow\infty}).
}
Thus, in the large $N_c$ limit, the hyperfine splittings are
\eqn\lnsplit{\eqalign{
\psq-\pq&={\kappa\over2m_Qm_q},\cr
\ssq-\sq&={3\kappa'\over4m_Qm_qN_c},\cr
(\sq)_{\rm avg}-\lq&= {\kappa'\over 2 N_c} \Delta \CH_\ell.\cr
}}
The  $P^*_Q-P_Q$ mass difference is of
order $1/m_Q$, the \split\ mass difference is of order
$(1/N_c)(1/m_Q)$, and the $(\sq)_{\rm avg}-\lq$ mass difference
is of order $1/N_c$.

The \split\ and $(\sq)_{\rm avg}-\lq$ mass differences
can be computed in the Skyrme model, in which baryons
containing a heavy quark are treated as soliton--heavy-meson bound
states. The formul\ae\ that are needed from earlier work are summarized
here; the details can be found in Refs.~\jmw\glm. The $P_Q$ and $P^*_Q$
states are combined into a heavy meson field
\eqn\hfield{
H_a = {(1 + \slash v)\over 2} \left[ P^*_{a\mu}\gamma^\mu - P_a
\gamma_5\right],
}
where $a=1,2$ denotes the $Q\bar u$ and $Q\bar d$ states,
and $v^\mu$ is the heavy quark four-velocity.
The transformation rule for the $H$ field under
$SU(2)_Q$ heavy quark spin symmetry is
\eqn\hsymmetry{
H_a\rightarrow S H_a,
}
where $S\in SU(2)_Q$, and the transformation rule under chiral $\twolr$ is
\eqn\hchiral{
H_a\rightarrow \left( H R^\dagger\right)_a,
}
where we use the primed basis for the $H$ fields defined in Ref.~\jmw.
The Goldstone boson fields have
the $\twolr$ transformation law
\eqn\sigmatrans{
\Sigma(x) \rightarrow L\ \Sigma(x)\ R^\dagger.
}
The $\Sigma$ field can be written in terms of the pion fields as
\eqn\sigmapion{
\Sigma(x) = e^{2i M /f},
}
where $M$ is the Goldstone-boson matrix
\eqn\Mdef{
M = \left[\matrix{
\pi^0/\sqrt2 & \pi^+ \cr
\noalign{\smallskip}
\pi^- & -\pi^0/\sqrt2 \cr
}\right],
}
and the pion decay constant $f\approx 132$~MeV.
The soliton solution of the $\twolr$ chiral lagrangian is
\eqn\skyrmesoln{
\Sigma = A\ \Sigma_0\ A^{-1} ,
}
where
\eqn\skyrmezero{
\Sigma_0 = \exp\left[ i F(\abs{\vec x})\ \hat x \cdot \vec \tau\right],
}
and $A\in SU(2)$ is the collective coordinate associated with isospin
transformations of the soliton solution $\Sigma_0$. The radial shape function
$F(r)$ satisfies $F(0)=-\pi$ and $F(\infty)=0$ for a soliton with baryon number
one. In the quantum theory, baryons have wavefunctions
that are functions of the matrix $A\in SU(2)$. The wavefunctions are
\eqn\wavefunction{
\psi_{R a m}(A)=(-1)^{R+m}\sqrt {\dim R}\ D^{*(R)}_{a\ -m}(A),
}
for a state with isospin $I=R$, $I_3=a$, $J=R$, and $J_3=m$. The matrices
$D^{(R)}$ are the representation matrices for $SU(2)$, and we have normalized
the measure on the $SU(2)$ group so that
\eqn\measure{
\int_{SU(2)} dA =1.
}
In QCD, the only soliton states in the theory are those with $2I=2J={\rm odd}$.

In the large $N_c$ limit, the soliton is very heavy, and the semiclassical
approximation is valid. Each time derivative is suppressed by a factor
of $1/N_c$, and can be neglected to
leading order in $N_c$. The interaction hamiltonian to leading order in
$N_c$ and in the spatial derivative expansion is
\eqn\intham{\eqalign{
\CH_I =&
- {ig\over 2} \int d^3 \vec x \
\Tr \bar H H \gamma^\mu \gamma_5
[\Sigma^{\dagger} \partial_\mu \Sigma] ,\cr
\approx&- {ig\over 2} \int d^3 \vec x \
\Tr \bar H H \gamma^j \gamma_5
[\Sigma^{\dagger} \partial_j \Sigma]  ,
}}
with $\Sigma$ given by Eq.~\skyrmesoln. In the limit that the $H$ field is very
heavy, the interaction energy is determined by the value of Eq.~\intham\ with
the $H$ field at the origin \jmw. Using the expansion
\eqn\fexp{
F(r) = F(0) + r F'(0) + \ldots =  -\pi + r F'(0) + \ldots,
}
in Eq.~\intham\ gives
the interaction hamiltonian \jmw\glm
\eqn\nint{\eqalign{
\CH_I &={g F'(0)\over 4}\int d^3 \vec x \
\Tr \bar H H \gamma^j \gamma_5 \tau^k \
\Tr A \tau^j A^{-1} \tau^k,\cr
&={g F'(0)}\ I^k_H\ S^j_{\ell H}\
\Tr A \tau^j A^{-1} \tau^k,\cr
}}
where $I_H$ and $S_{\ell H}$ are the isospin and spin of the light
degrees of freedom of $H$. The $\sq$, $\ssq$ multiplet has $I=1$ and
$s_\ell=1$. The soliton states have $I=s_\ell=1/2,\ 3/2,\ \ldots$, and the
light degrees of freedom of $H$ have $I=s_\ell=1/2$.
There are two states in the soliton-meson spectrum with $I=s_\ell=1$,
obtained by tensoring the light degrees of
freedom of the $H$ particle with either the $N$ ($I=s_\ell=1/2$)
or the $\Delta$ ($I=s_\ell=3/2$) soliton
states. These states will be denoted by
$\ket{1\ 1;N}$ and $\ket{1\ 1;\Delta}$,
where the first label refers to isospin and the second to the spin of
the light degrees of freedom.
The interaction hamiltonian Eq.~\nint\ in the $\ket{1\ 1 ;N,\Delta}$
basis can be evaluated explicitly in terms of Clebsch-Gordan
coefficients \glm,
\eqn\twostate{
\CH_I = -{ g F'(0)\over 6} \pmatrix{1&4\sqrt2\cr4\sqrt2&5\cr}.
}
The two $\ket{1\ 1}$ states which are the eigenstates of $\CH_I$
in Eq.~\twostate\ are
\eqn\twostates{\eqalign{
\kket{1\ 1}_0 &= \sqrt{\bfrac13}\ \ket{1\ 1; N} + \sqrt{\bfrac23}\ \ket{1\ 1;
\Delta},\cr\noalign{\smallskip}
\kket{1\ 1}_1 &= \sqrt{\bfrac23}\ \ket{1\ 1; N} - \sqrt{\bfrac13}\ \ket{1\ 1;
\Delta},
}}
with energies $-3 g F'(0)/2$ and $g F'(0)/2$, respectively. The state
$\kket{1\ 1}_0$ when combined with the heavy quark in $H$ is the $\sq$,
$\ssq$
multiplet, and $\kket{1\ 1}_1$ combined with the heavy quark is
an unbound exotic multiplet. The subscript
on the $\kket{\ }$ states is the eigenvalue of $K=I+S_\ell$; the
significance of $K$ is
discussed in detail in Ref.~\glm. The $\lq$ has $I=s_\ell=0$. The
only state in the soliton-meson spectrum with the correct quantum
numbers is in the nucleon sector, $\ket{0\ 0;N}$. The interaction energy
of this state due to the hamiltonian Eq.~\nint\ is $-3gF'(0)/2$ \jmw. The
state $\ket{0\ 0;N}$ has $K=0$, and will also be denoted by $\kket{0\ 0}_0$.

The $\ssq,\ \sq$ and $\lq$
states are obtained by combining the spin of the light
degrees of freedom with the spin of the heavy quark. Since  spin
states with different values of $s_3$ all
have the same mass, only the $\Sigma_Q^*$ state with $s_3=3/2$,
the $\Sigma_Q$ state with $s_3=1/2$, and the $\lq$ state with
$s_{3}=1/2$
are needed to determine the energies,
\eqn\sstar{\eqalign{
\ket{\Sigma_Q^*,\ \frac32} &= \kket{1\ 1\ 1}_0\ket{\uparrow}_Q\cr
\noalign{\smallskip}
\ket{\Sigma_Q,\ \frac12} &= \sqrt\bfrac23\kket{1\ 1\ 1}_0\ket{\downarrow}_Q
-\sqrt\bfrac13\kket{1\ 1\ 0}_0\ket{\uparrow}_Q\cr
\noalign{\smallskip}
\ket{\lq,\ \frac12} &= \kket{0\ 0\ 0}_0\ket{\uparrow}_Q.
}}
In Eq.~\sstar,
the state $\ket{\ }_Q$ denotes the spin state of the heavy quark, and
$\kket{I\ s_\ell\ m}_0$ represents the state of the light degrees of
freedom $\kket{I\ s_\ell}_0$
with $s_{\ell3}=m$. The coefficients in Eq.~\sstar\ are the
Clebsch-Gordan coefficients for $1\otimes 1/2\rightarrow 3/2$,
$1\otimes 1/2\rightarrow 1/2$, and $0\otimes 1/2\rightarrow 1/2$,
respectively.

The leading term in the $1/N_c$ and spatial
derivative expansion that splits the $N, \Delta$
degeneracy is the kinetic term of the soliton, which has two time
derivatives and no space derivatives, and is of order $1/N_c$.
Including this term shifts the energy of
the $\ket{1\ 1;\Delta}$ states, and changes the
interaction hamiltonian Eq.~\twostate\ to
\eqn\newint{
\CH_I = -{ g F'(0)\over 6} \pmatrix{1&4\sqrt2\cr4\sqrt2&5\cr} +
\pmatrix{0&0\cr0&\Delta M\cr},
}
where $\Delta M$ is the $\Delta-N$ mass difference, and is of order
$1/N_c$. The eigenstate of Eq.~\newint\ that (combined with the heavy
quark states) produces the $\sq$, $\ssq$ multiplet
and reduces to $\kket{1\ 1}_0$ in the limit $\Delta M\rightarrow 0$ is
\eqn\lower{
\kket{1\ 1}_\epsilon = a \ket{1\ 1; N} + b \ket{1\ 1; \Delta}
=\kket{1\ 1}_0 + \epsilon \kket{1 \ 1}_1,
}
where
\eqn\coefs{
a = \sqrt\bfrac13 + \sqrt\bfrac23\ \epsilon,\ \ \ b=\sqrt\bfrac23 -
\sqrt\bfrac13\ \epsilon ,\ \ \
\epsilon={\Delta M\over 3 \sqrt2 g F'(0)} ,
}
to first order in $\Delta M$. The $\sq,\ \ssq$ states to order $1/N_c$
are given by Eq.~\sstar\ with $\kket{1\ 1}_0$ replaced by $\kket{1\
1}_\epsilon$. Since the $\lq$ state $\kket{0\ 0}_0$ only contains the
soliton in the nucleon sector, its interaction energy $-3gF'(0)/2$
is unaffected by the $\Delta-N$ mass difference.

The soliton-meson
states $\ket{1\ 1;N}$ and $\ket{1\ 1;\Delta}$ which appear in
Eq.~\lower\ can
be written explicitly in terms of the soliton states and the spin
of the light degrees of freedom of $H$,
\eqn\explicit{\eqalign{
\ket{1\ 1\ 1;N} &= \ket{\frac12}_N\ket{\uparrow}_\ell ,\cr
\noalign{\smallskip}
\ket{1\ 1\ 0;N} &= \sqrt\bfrac12\ket{\frac12}_N\ket{\downarrow}_\ell
+\sqrt\bfrac12\ket{-\frac12}_N\ket{\uparrow}_\ell ,\cr
\noalign{\smallskip}
\ket{1\ 1\ 1;\Delta} & = \sqrt\bfrac34\ket{\frac32}_\Delta\ket{\downarrow}_\ell
-\sqrt\bfrac14\ket{\frac12}_\Delta\ket{\uparrow}_\ell ,\cr
\noalign{\smallskip}
\ket{1\ 1\ 0;\Delta} & = \sqrt\bfrac12\ket{\frac12}_\Delta\ket{\downarrow}_\ell
-\sqrt\bfrac12\ket{-\frac12}_\Delta\ket{\uparrow}_\ell ,\cr
\noalign{\smallskip}
\ket{0\ 0\ 0;N} &= \sqrt\bfrac12\ket{\frac12}_N\ket{\downarrow}_\ell
-\sqrt\bfrac12\ket{-\frac12}_N\ket{\uparrow}_\ell ,\cr
}}
where $\ket{\ }_\ell$ is the spin state of the light degrees of freedom
in the heavy meson,
$\ket{m}_{N,\Delta}$ is the soliton state in the $N$ or $\Delta$ sector
with $s_{\ell3}=m$, and $\ket{I\ s_\ell\ m;N,\Delta}$ is the bound state
$\ket{I\ s_\ell;N, \Delta}$ with $s_{\ell3}=m$. The coefficients in
Eq.~\explicit\ are the Clebsch-Gordan coefficients for $1/2\otimes
1/2\rightarrow 1$, $3/2\otimes 1/2\rightarrow 1$ and $1/2\otimes
1/2\rightarrow 0$ for the $\ket{1\ 1;N}$, $\ket{1\ 1;\Delta}$ and
$\ket{0\ 0;N}$ states, respectively.
Finally, the tensor product of the light
degrees of freedom and the heavy quark in $H$ can be reexpressed in
terms of $P$ and $P^*$ mesons,
\eqn\reexpress{\eqalign{
\ket{\uparrow}_\ell\ket{\uparrow}_Q&=\ket{\psq,1} ,\cr
\noalign{\smallskip}
\ket{\downarrow}_\ell\ket{\downarrow}_Q&=\ket{\psq ,-1} ,\cr
\noalign{\smallskip}
\ket{\uparrow}_\ell\ket{\downarrow}_Q&=\sqrt\bfrac12\ket{\pq}+\sqrt\bfrac12
\ket{\psq,0} ,\cr
\noalign{\smallskip}
\ket{\downarrow}_\ell\ket{\uparrow}_Q&=-\sqrt\bfrac12\ket{\pq}+\sqrt\bfrac12
\ket{\psq,0} ,\cr
}}
where $\ket{\psq,m}$ is the $\psq$ meson with $s_3=m$. Combining
Eqs.~\sstar--\reexpress, gives
\eqn\fullstate{\eqalign{
\ket{\ssq,\ \frac32} &=  -\sqrt\bfrac38\ b\ \ket{\frac32}_\Delta\ket{\pq} -
\bfrac12\ b\ \ket{\frac12}_\Delta\ket{\psq,1} + a
\ \ket{\frac12}_N\ket{\psq,1}\cr
&\qquad\qquad\qquad+ \sqrt\bfrac38\ b\ \ket{\frac32}_\Delta \ket{\psq,0} ,\cr
\noalign{\smallskip}
\ket{\sq,\ \frac12} &=  \sqrt\bfrac34\ a\ \ket{\frac12}_N\ket{\pq} +
\sqrt\bfrac12\ b\ \ket{\frac32}_\Delta\ket{\psq,-1} +
\sqrt\bfrac16\ b\ \ket{-\frac12}_\Delta\ket{\psq,1}\cr
&-  \sqrt\bfrac16\ a\  \ket{-\frac12}_N \ket{\psq,1}
-\sqrt\bfrac13\ b\ \ket{\frac12}_\Delta\ket{\psq,0} + \sqrt\bfrac1{12}\ a\
\ket{\frac12}_N\ket{\psq,0},\cr
\noalign{\smallskip}
\ket{\lq,\ \frac12} &= \bfrac12\ \ket{\frac12}_N \ket{\psq,0} -
\bfrac12\ \ket{\frac12}_N \ket{\pq}-
\sqrt\bfrac12\ \ket{-\frac12}_N\ket{\psq,1}.\cr
}}

The leading operator in the derivative expansion which breaks the heavy
quark symmetry is
\eqn\msplit{
\CL_1={\lambda_2\over m_Q} \Tr \bar H \sigma^{\mu\nu} H \sigma_{\mu\nu}
}
which yields a $P^*_Q-P_Q$ mass difference $-8\lambda_2/m_Q$.
The $\ssq,\, \sq,\, \lq$  masses including Eq.~\msplit\
can be read off directly from the expressions for the states in
Eq.~\fullstate,
\eqn\energies{\eqalign{
\Sigma_Q^*&= -{3 g F'(0)\over2} + a^2 N + b^2 \Delta + {3 b^2\over
8} \pq + \left(a^2+ {5 b^2\over8}\right) \psq,\cr
\noalign{\smallskip}
\Sigma_Q&=- {3 g F'(0)\over2} + a^2 N + b^2 \Delta + {3 a^2\over
4} \pq + \left({a^2\over4}+b^2\right) \psq,\cr
\noalign{\smallskip}
\lq&= -{3 g F'(0)\over2} + N + \bfrac14 \pq + \bfrac34 \psq,\cr
}}
where the coefficient $a^2$ is the probability that $\ssq$ contains a
nucleon, $3b^2/8$ is the probability that $\ssq$ contains a $\pq$, \etc\
The mass differences are
\eqn\mdiff{\eqalign{
\noalign{\medskip}
\ssq-\sq&={3\left[2 a^2 - b^2\right]
\over 8}\left(\psq-\pq\right),\cr
(\sq)_{\rm avg}-\lq&=b^2\left(\Delta-N\right),\cr
\noalign{\smallskip}
}}
using $a^2+b^2=1$.
Substituting the values of $a$ and $b$ from Eq.~\coefs\ gives
\eqn\mdifftwo{\eqalign{
\noalign{\smallskip}
\ssq-\sq&={\left(\Delta-N\right)\left(\psq-\pq\right)
\over 4g F'(0)},\cr
(\sq)_{\rm avg}-\lq&={2\over3}\left(\Delta-N\right).\cr
\noalign{\smallskip}
}}
Note that
the mass splittings have the correct dependence on $m_Q$ and $N_c$.
The
$\psq-\pq$ mass difference is of order $1/m_Q$, and the $\Delta-N$ mass
difference is of order $1/N_c$. This implies that the \split\ mass difference
is of order $1/(m_QN_c)$ and the $(\sq)_{\rm avg}-\lq$ mass difference
is of order
$1/N_c$, as expected from consideration of the constituent quark model.

The value of $gF'(0)$ is obtained from the $\Lambda_c$ mass in
Eq.~\energies\ to be 418~MeV, so that
\eqn\values{\eqalign{
\Sigma^*_c-\Sigma_c=25\ \MeV,\qquad \Sigma^*_b-\Sigma_b=8\
\MeV,\cr
(\Sigma_c^*)_{\rm avg}-\Lambda_c=
(\Sigma_b^*)_{\rm avg}-\Lambda_b=195\ \MeV.\quad\cr
}}
This gives the predictions
\eqn\predict{\eqalign{
({\rm input})\ \ \Lambda_c=2285\ \MeV\,\qquad
\Lambda_b=5625\ \MeV,\cr
\Sigma_c=2464\ \MeV,\qquad
\Sigma_b=5816\ \MeV,\cr
\Sigma_c^*=2489\ \MeV,\qquad
\Sigma_b^*=5824\ \MeV,\cr
}}
to be compared with the experimental values $\Sigma_c=2453$ MeV,
and $\Lambda_b=5641\pm40$ MeV. The predictions are in good experiment
with experiment. Note that the typical magnitude of the
$\Sigma_c-\Lambda_c$ splitting is of order the $\Delta-N$ splitting,
which is $\approx300$~MeV. The Skyrme model predicts the value of this
splitting to within $\approx10$~MeV.
The masses of the charm and bottom
baryons  have also been computed in the constituent quark model
\ref\capstick{S. Capstick and N. Isgur, \physrev{D34}{1986}{2809}}.
The quark model model prediction of the \split\ splitting is about
twice as big as the Skyrme model prediction.

The hyperfine splittings in Eq.~\mdifftwo\ have been computed to leading
order in the $1/N_c$ expansion, and to leading order in the spatial
derivative expansion, so they are not true large $N_c$ predictions. For
example, the operator
\eqn\pcoup{
\CL_2 = {\lambda_\pi\over m_Q}\Tr\bar H \gamma^\mu\gamma_5 H \,
\Sigma^\dagger
\partial_\mu \Sigma,
}
violates the heavy quark symmetry, and destroys the equality between the
$\psq\psq\pi$ and $\psq\pq\pi$ coupling constants. This operator will
produce a \split\ mass difference proportional to
$(\lambda_\pi/m_Q)(\Delta-N)$. The operator $\CL_2$ has one space
derivative, whereas $\CL_1$ in Eq.~\msplit\ has no space derivative.
If the derivative expansion is valid, the contribution of $\CL_2$ to the
\split\ mass difference should be smaller than the $\CL_1$ contribution that
we have computed.

Some general features of the results in this paper can be easily
understood on the basis of $K$ invariance. In the large $N_c$ limit,
baryon states can be classified by the quantum number $K$, and the $\lq$
and $\sq,\,\ssq$ states all have $K=0$. At order $1/N_c$ the
$\sq,\,\ssq$ states are a superposition of a $K=0$ state with
amplitude $N_c^0$ and a $K=1$ state with amplitude $1/N_c$. The $\lq$
state remains a pure $K=0$ state because there is no $K=1$ state with the
correct quantum numbers to mix with the $\lq$. The \split\ mass
difference violates the heavy quark symmetry, and therefore must come
from an operator in the effective hamiltonian that involves $S_Q$, the
spin of the heavy quark. By rotational invariance, $S_Q$ must be dotted
into the spin of the light degrees of freedom of either $H$ or the
soliton. (The orbital angular momentum contributions are suppressed by
$1/N_c$ or by an additional $1/m_Q$.) The operator that breaks the heavy
quark symmetry must therefore transform as a $K=1$ irreducible
tensor operator, since $K=I+S_\ell$. A $K=1$ tensor operator cannot have
a matrix element between two $K=0$ states. Thus the operator that breaks
the heavy quark symmetry does not affect the $\lq$ energy, and only
contributes to the $\sq,\, \ssq$ energy at order $1/N_c$.
The $1/m_Q$ terms such as Eq.~\msplit\ are of leading order in
$1/N_c$ because they contain no time derivatives. The $1/N_c$
$K$-violation arises from the $\Delta-N$ mass difference in the soliton
sector. There also are terms that simultaneously violate the heavy quark
symmetry and the $K$-symmetry, and are of order $1/(m_QN_c)$ because they
contain one time derivative, \eg
\eqn\kviolone{
\CL_3 = {1\over m_Q}
\epsilon_{\mu\nu\alpha\beta}\Tr\bar H \,\gamma^\mu
\gamma_5 H \,\Sigma^\dagger \partial^\nu \Sigma\partial^\alpha
\Sigma^\dagger\partial^\beta\Sigma\left(\partial^2\Sigma^\dagger\Sigma-
\Sigma^\dagger\partial^2\Sigma\right),
}
or
\eqn\kviotwo{
\CL_4={1\over m_Q}
\Tr\bar H \,\gamma^\mu \gamma_5 \,H \,\gamma_\mu\gamma_5 \,
\Sigma^\dagger\left(v\cdot\partial\right)\Sigma
\left(\partial^2\Sigma^\dagger\Sigma-
\Sigma^\dagger\partial^2\Sigma\right).
}
Both terms start at one time derivative and four space derivatives, and
are expected to be small. ($\CL_4$ naively starts at two space
derivatives, but $\Sigma^\dagger d\Sigma/dt$ vanishes at the origin
where the $H$ particle is bound, which brings in an extra suppression factor.)

\bigskip
\centerline{\bf Acknowledgements}
We would like to thank the Fermilab theory group for
hospitality.
This work was supported in part by DOE grant \doe,
and by a NSF Presidential Young Investigator award \pyiam.
\bigskip

\listrefs
\bye